\documentclass{iopjournal}

\usepackage{ragged2e}
\usepackage{amssymb}
\usepackage{amsmath}
\usepackage{pgfplots}

\newcommand\bb{\boldsymbol{b}}
\newcommand\bB{\boldsymbol{B}}

\pgfplotsset{compat=1.18}

\begin{document}
\justifying

% \articletype{Paper} %	 e.g. Paper, Letter, Topical Review...

\title{Suppressed Stiffness of energetic particle transport due to thermal plasma nonlinearity in tokamak plasmas}
\author{Guo Meng$^{1}$, Zhixin Lu$^{1}$, Gengxian Li$^{1}$ }

\affil{$^1${Max Planck Institute for Plasma Physics, Boltzmannstr. 2,  Garching, 85748, Germany}}

%\email{zhixin.lu@ipp.mpg.de}


\begin{abstract}
Energetic particle (EP) transport stiffness plays a crucial role in determining EP confinement in tokamak plasmas. This work investigates the impact of nonlinear thermal plasma dynamics on EP transport using global gyrokinetic simulations with the TRIMEG code. Simulations are performed for the ITPA toroidal Alfvén eigenmode (TAE) benchmark, and extended to beta-induced Alfvén eigenmodes (BAE) and reversed shear Alfvén eigenmodes (RSAE). Two models are compared: a fully nonlinear treatment of all species and a reduced model with only nonlinear EP.
For the TAE simulations, while the linear properties of the instability are identical in both cases, significant differences arise in the nonlinear regime. The inclusion of thermal plasma nonlinearity leads to a reduced saturation level following an overshoot phase and modifies the radial mode structure, including mode broadening and poloidal harmonic splitting. These changes significantly affect EP transport.
In particular, the dependence of the EP flux on the EP drive becomes weaker when nonlinear thermal plasma dynamics are included. The saturation level changes from an approximately quadratic scaling to a weaker, nearly linear dependence, leading to a reduced scaling of EP flux with the EP gradient from  quartic to  quadratic. Zonal flow generation is observed but found to play a minor role in regulating the instability. Without accounting for thermal nonlinearity, the EP flux can be overestimated by an order of magnitude.  
In addition to the TAE, the BAE and RSAE have also been simulated, demonstrating the similar effects of the thermal plasma nonlinearity on mitigating the saturation level. 
These results demonstrate that nonlinear thermal plasma effects provide an important feedback mechanism that reduces EP transport stiffness and regulates Alfvénic mode saturation. This feature is therefore essential for predictive modeling of energetic particle confinement in burning plasmas such as ITER.
\end{abstract}

\section{Introduction}
Energetic particles (EPs) play an important role in the dynamics of magnetically confined plasmas. In burning plasma devices such as ITER, fusion-born alpha particles are expected to provide a substantial fraction of the plasma heating power and must remain well confined to sustain the burning state. However, energetic particles can resonantly interact with shear Alfv\'en waves and destabilize a variety of collective modes, leading to redistribution and enhanced EP transport. 

Energetic particle-driven instabilities have been widely studied in tokamak plasmas. Energetic particles can excite several types of Alfvénic modes, including toroidal Alfvén eigenmodes (TAEs), reversed-shear Alfvén eigenmodes (RSAEs), and energetic particle modes (EPMs). The theoretical framework describing these processes has been extensively developed within kinetic and gyrokinetic theory, as summarized in the reviews by Chen and Zonca \cite{chen2016physics}. Nonlinear wave-particle interactions and phase-space dynamics have also been studied in the work of Breizman, Berk and collaborators \cite{berk1992scenarios}, while the impact of EP-driven instabilities on confinement has been reviewed by Gorelenkov \cite{gorelenkov2014energetic}. Critical-gradient behavior in Alfvén eigenmode-induced fast-ion transport has been reported, showing that EP transport suddenly becomes stiff above a critical threshold in the presence of many overlapping small-amplitude Alfvén eigenmodes \cite{collins2016observation}. 

In realistic tokamak plasmas, the nonlinear evolution of EP-driven modes is influenced by the collective dynamics of the background thermal plasma \cite{vlad2025state}. However, in many numerical studies, the thermal ions and electrons are treated using simplified or linearized models in order to reduce computational cost. While such approaches capture important aspects of EP-driven instabilities, they may miss nonlinear feedback mechanisms associated with the self-consistent evolution of the thermal plasma. Recent theoretical work has considered the effects of the thermal ions and electrons by constructing the theoretical framework of the Reversed Shear Alfv\'enic Eigenmode-Zonal flow-Kinetic Alfv\`en Wave \cite{ma2025zonal}. The simulation and theoretical work have been carried out to address the nonlinear thermal plasma effect on the TAE saturation, where the nonlinear modification of the curvature coupling term has been identified as the key ingredient of the nonlinear saturation of the EP-driven TAEs \cite{chen2026nonlinear}. In the studies of the Kinetic ballooning mode (KBM), the parallel nonlinearity of the electrons is found to play a key role in the saturation \cite{chen2025saturation}. 

In this work, we investigate the effect of nonlinear thermal plasma dynamics on energetic particle transport using global gyrokinetic simulations with the TRIMEG code \cite{lu2019development,lu2021development,lu2025trimeg}. Simulations are performed for the ITPA TAE benchmark case \cite{konies2018benchmark} and compare two models: one in which only energetic particles evolve nonlinearly, and another in which all plasma species are treated nonlinearly. The results show that nonlinear thermal plasma dynamics can significantly modify the mode structure and lead to a stiff dependence of energetic particle transport. In addition, simulations of beta-induced Alfvén eigenmodes (BAEs) and reversed-shear Alfvén eigenmodes (RSAEs) are performed by modifying the safety factor ($q$) profiles to further identify and  illustrate the underlying mechanisms more generally.

The remainder of this paper is organized as follows. Section~\ref{sec:models} describes the gyrokinetic model and simulation parameters. Section~\ref{sec:mode_property} discusses the linear and nonlinear properties of the modes. Section~\ref{sec:transport} analyzes the saturation level and transport properties. Phase-space dynamics are presented in Section~\ref{sec:phase_space}, followed by the discussion of the effect of zonal flow in Section~\ref{sec:zonal}. The stiffness of the saturation and the transport level for varying EP driven are presented in Section~\ref{sec:stiffness}. The results of BAE and RSAE simulation are discussed in Sections~\ref{sec:BAE} and \ref{sec:RSAE}, followed by conclusions in Section~\ref{sec:summary}.

\section{Simulation model and parameters}
\label{sec:models}
\subsection{Simulation model in TRIMEG code}
We use TRIMEG in this work for the gyrokinetic particle simulations. The mixed variable-pullback scheme has been implemented \cite{mishchenko2014pullback,hatzky2019reduction,lanti2020orb5,lu2023full}. 
Using the mixed variable scheme, the parallel component $\delta A_\|$ of the perturbed magnetic potential is decomposed into a symplectic part and a Hamiltonian part \cite{mishchenko2014pullback},
\begin{equation}
\label{eq:AsAh}
    \delta A_\| =\delta A_\|^{\rm{s}} + \delta A_\|^{\rm{h}} \;\;.
\end{equation}
The shifted parallel velocity coordinate of the gyrocenter $u_\|$ is defined as 
\begin{equation}
    u_\|=v_\|+\frac{q_s}{m_s}\langle\delta A_\|^{\rm{h}}\rangle\;\;,
\end{equation}
where $v_\|$ is the parallel velocity, $q_s$ and $m_s$ are the charge and mass of species $s$, respectively, the subscript ``$s$'' represents the different particle species, and $\langle\ldots\rangle$ indicates the gyro average.

The gyrocenter equations of motion are consistent with previous work \cite{mishchenko2014pullback,hatzky2019reduction,lanti2020orb5,mishchenko2023global,kleiber2024euterpe}, 
\begin{eqnarray}
\label{eq:dR0dt}
 \dot{\boldsymbol R}_0 
  &=& u_\| {\boldsymbol b}^*_0 + \frac{m\mu}{qB^*_\|} {\boldsymbol b}\times\nabla B \;\;, 
  \\
\label{eq:du0dt}
  \dot u_{\|,0}
  &=& -\mu {\boldsymbol b}^*_0\cdot \nabla B \;\;,
  \\
  \label{eq:dR1dt}
  \delta\dot{\boldsymbol R}
  &=& \frac{{\boldsymbol b}}{B^*_\|}\times \nabla \langle \delta\Phi -u_\| \delta A_\|\rangle 
  -\frac{q_s}{m_s}\langle\delta A^{\rm h}_\|\rangle {\boldsymbol b}^*\;\;, 
  \\
  \delta \dot u_\|
  &=&  -\frac{q_s}{m_s} \left({\boldsymbol b}^*\cdot\nabla\langle\delta\Phi-u_\|\delta A^{\rm{h}}_\|\rangle +\partial_t\langle\delta A_\|^{\rm{s}}\rangle \right) \nonumber\\
  \label{eq:du1dt}
  &&-\frac{\mu}{B^*_\|}{\boldsymbol b}\times\nabla B\cdot\nabla\langle\delta A_\|^{\rm{s}}\rangle \;\;,  
\end{eqnarray}
where the magnetic moment $\mu=v_\perp^2/(2B)$, 
${\boldsymbol b}^*={\boldsymbol b}_0^*+\nabla\times({\boldsymbol b}\langle\delta A_\|^{\rm s}\rangle)/B_\|^*
\approx{\boldsymbol b}_0^*+\nabla\langle\delta A_\|^{\rm s}\rangle\times{\boldsymbol b}/B_\|^*$, ${\boldsymbol b}^*_0={\boldsymbol b}+(m_s/q_s)u_\|\nabla\times{\boldsymbol b}/B_\|^*$, ${\boldsymbol b}={\boldsymbol B}/B$, $\boldsymbol{B}$ is the equilibrium magnetic field, $B_\|^*=B+(m_s/q_s)u_\|{\boldsymbol b}\cdot(\nabla\times{\boldsymbol b})$. The modified magnetic field is 
$\bB^*=\bB+\bB_1+ (m_s/q_s) u_\parallel \nabla \times \bb$.

The distribution function is decomposed into the equilibrium one, and the perturbed one $f=f_0+\delta f$, and $\delta f$ is solved along the particle trajectory using the standard procedure \cite{lanti2020orb5,hatzky2019reduction,mishchenko2014pullback,lu2023full}. In particle simulations, $\delta f$ is represented by the numerical markers
$\delta f=C_{\rm P2G}\sum_{p=1,N_{\rm mark}} ({w_p}/{J_z}) \delta({\bf z}-{\bf z}_p)$, $C_{\rm P2G}={\langle n\rangle_V V_{\rm tot}}/{N_{\rm mark}}$
where $N_{\rm mark}$ is the marker number, $ \langle\ldots\rangle_V$ indicates the volume average, and $V_{\rm tot}$ is the total volume. $\delta f$ is solved along the particle trajectory,
\begin{eqnarray}
\label{eq:dfdt}
    \frac{d\delta f}{dt}\equiv \frac{\partial\delta f}{\partial t}
    +\left(\dot{\boldsymbol R}_0 +\delta\dot{\boldsymbol R}\right)\cdot\nabla \delta f
    +\left(\dot u_{\|,0}+ \delta \dot u_\|\right)\frac{\partial \delta f}{\partial u_\|}
    =-\delta\dot{\boldsymbol R}\cdot\nabla f_0
    -\delta \dot u_\|\frac{\partial f_0}{\partial v_\|} \;\;,
\end{eqnarray}
where $f_0$ is chosen as a Maxwellian distribution $f_M$, and $f_M$ is assumed to be a steady state solution when the perturbed field variables are absent to eliminate the neoclassical physics. 

The linearized quasi-neutrality equation in the long-wavelength approximation is as follows, 
\begin{equation}
\label{eq:poisson0}
    -\nabla\cdot\left( \sum_s\frac{q_s n_{0s}}{B\omega_{{\rm c}s}} \nabla_\perp\delta\Phi \right) = \sum_s q_s \delta n_{s,v} \;\;,
\end{equation}
where the gyrocenter density $\delta n_{s,v}$ is calculated using $\delta f_s({\boldsymbol R},v_\|,\mu)$ (indicated as $\delta f_{s,v}$), namely, $\delta n_{s,v}({\boldsymbol{x} })=\int {\rm d}^6 z\delta f_{s,v}\delta(\boldsymbol{R}  + \boldsymbol{\rho} - \boldsymbol{x} )$. Here, $\boldsymbol{x} $ and $\boldsymbol{R} $ denote the particle position vector and gyrocenter position vector, respectively, and $\boldsymbol{\rho}$ represents the Larmor radius.
In Eq.~\eqref{eq:poisson0}, $\omega_{{\rm c}s}$ is the cyclotron frequency of species $s$, and in this work, we ignore the perturbed electron polarization density on the left-hand side. 
When the $\delta f$ scheme is adopted, $\delta f_{s,v}$ is obtained from $\delta f_{s,u}$ as follows, with the linear approximation of the pullback scheme,
\begin{eqnarray}
    & \delta f_{s,v} = \delta f_{s,u} +  \frac{q_s\left\langle\delta A^{\rm{h}}_{\|} \right\rangle}{m_s}\frac{\partial f_{0s}}{\partial v_\|}
    \xrightarrow[f_{0s}=f_{\rm{M}}]{\text{Maxwellian}}
     \delta f_{s,u} -  \frac{ v_\|}{T_s}  q_s\left\langle\delta A^{\rm{h}}_{\|} \right\rangle f_{0s}\;\;.
\end{eqnarray}

Amp\`ere's law in $v_\|$ space is 
$-\nabla^2_\perp\delta A_\| = \mu_0 \delta j_{\|,v}$
where $\delta j_{\|,v}({\boldsymbol{x} })=\sum_s q_s \int {\rm d}^6 z\delta f_{s,v}\delta(\boldsymbol{R}  + \boldsymbol{\rho} - \boldsymbol{x} )v_\|$. It is solved in the mixed variable space. 
Using an iterative scheme \cite{chen2007electromagnetic,mishchenko2017mitigation,hatzky2019reduction,lu2023full}, the asymptotic solution is expressed as $\delta A^{\rm{h}}_{\|}=\sum_{I=0}^\infty\delta A^{\rm{h}}_{\|,I}$,
where $|\delta A^{\rm{h}}_{\|,I+1}/\delta A^{\rm{h}}_{\|,I}|\ll1$ is assured by the fact that the analytical skin depth term is close to the exact one. 
Amp\`ere's law is solved iteratively as follows,
\begin{eqnarray}
\label{eq:ampere_h0}
    \left(\nabla^2_\perp-\sum_s\frac{1}{d_{s}^2}\right)\delta A_{\|,0}^{\rm{h}} 
    &=& -\nabla^2_\perp\delta A_{\|}^{\rm{s}} - \mu_0 \delta j_{\|} \;\;, \\
\label{eq:ampere_iterative}
    \left(\nabla^2_\perp-\sum_s\frac{1}{d_{s}^2}\right)\delta A_{\|,I}^{\rm{h}} 
    &=&-\sum_s\frac{1}{d_{s}^2}\delta A^{\rm{h}}_{\|,I-1} 
    + \sum_s\frac{1}{d_{s}^2} \overline{\langle\delta A_{\|,I-1}^{\rm{h}}\rangle}\;\;, \\
    \overline{\langle\delta A_{\|,I-1}^{\rm{h}}\rangle}
    &=&\frac{2}{n_0 v_{\rm{t}s}^2}\int \mathrm{d}z^6 v_\|^2 f_{0s}  \langle\delta A^{\rm{h}}_{\|,I-1} \rangle\delta(\boldsymbol{R}  + \boldsymbol{\rho} - \boldsymbol{x} )  \;\;\text{ for $\delta f$}\;\;, \\
\label{eq:A2ndavg_fullf}
    \overline{\langle\delta A_{\|,I-1}^{\rm{h}}\rangle}
    &=&\frac{1}{n_0}\int \mathrm{d}z^6 f_{s,v}  \langle\delta A^{\rm{h}}_{\|,I-1} \rangle \delta(\boldsymbol{R}  + \boldsymbol{\rho} - \boldsymbol{x} ) \;\;\text{ for full $f$  \;\;,}
\end{eqnarray}
where  $I=1,2,3,\ldots$ and details of the $\delta f$ model and full $f$ model can be found in the previous works \cite{hatzky2019reduction,lu2023full}, $v_{th,s}=\sqrt{2T_s/m_s}$, $T_s$ is the temperature, $d_s=c/\omega_{\mathrm{p}s}$ is the skin depth, $\omega_{\mathrm{p}s}=\sqrt{n_sq_s^2/(m_s\varepsilon_0)}$ is the plasma frequency of species "s".

The pullback scheme for the $\delta f$ method can be found in the previous work \cite{mishchenko2014pullback} as follows,
\begin{align}
\label{eq:pullback_A}
    & \delta A^{\rm{s}}_{\|,\rm{new}} = \delta A^{\rm{s}}_{\|,\rm{old}} + \delta A^{\rm{h}}_{\|,\rm{old}}  \;\;, \\
\label{eq:pullback_v}
    & u_{\|,\rm{new}} = u_{\|,\rm{old}} - \frac{q_s}{m_s} \left\langle\delta A^{\rm{h}}_{\|,\rm{old}} \right\rangle  \;\;, \\
\label{eq:pullback_df}
    & \delta f_{\rm{new}} = \delta f_{\rm{old}} +  \frac{q_s\left\langle\delta A^{\rm{h}}_{\|,\rm{old}} \right\rangle}{m_s}\frac{\partial f_{0s}}{\partial v_\|}
    \xrightarrow[f_{0s}=f_{M}]{\text{Maxwellian}}
     \delta f_{\rm{old}} -  \frac{ 2v_\|}{v_{\rm{t}s}^2}  \frac{q_s\left\langle\delta A^{\rm{h}}_{\|,\rm{old}} \right\rangle}{m_s} f_{0s} \;\;,
\end{align}
where variables with subscripts ``new'' and ``old'' refer to those after and before the pullback transformation, Eq.~(\ref{eq:pullback_df}) is the linearized equation for $\delta f$ pullback, which is from the general equation of the transformation for the distribution function $
    f_{\rm{old}} (u_{\| \rm{old}}) = f_{\rm{new}} \left(u_{\| \rm{new}} =u_{\| \rm{old}}- {q_s} \left\langle\delta A^{\rm{h}}_{\|,\rm{old}} \right\rangle/{m_s} \right)$. 
For the full $f$ scheme, only Eqs.~(\ref{eq:pullback_A}) and (\ref{eq:pullback_v}) are needed. 

\subsection{Parameters of the ITPA-TAE case}
\begin{figure}
 \centering
        \includegraphics[width=0.8\textwidth]{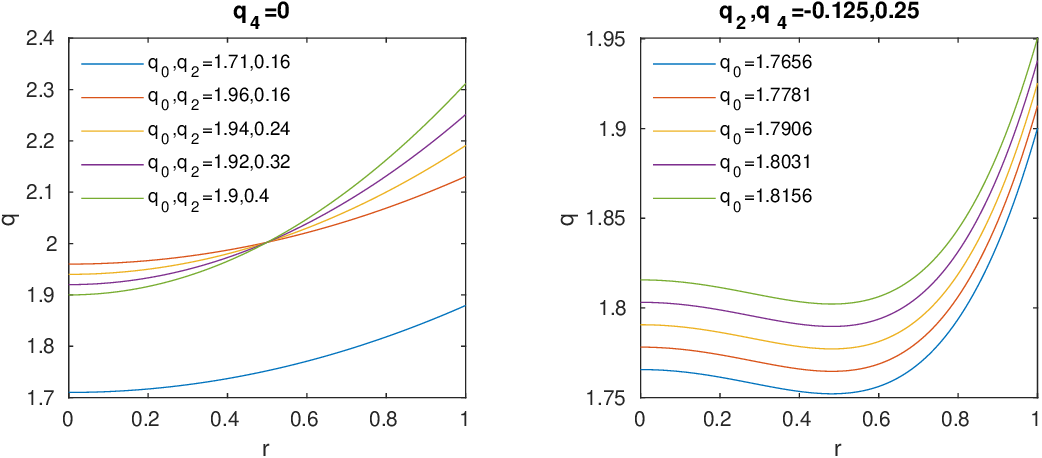}
 \caption{The $q$ profiles used in this work. Left: the $q$ profile for the simulations of the TAE (blue line) and BAE with different values of the local magnetic shear at $r_c=0.5$ (the other four lines). Right: the $q$ profile for the simulations of the RSAE with different values of $q_{\rm min}$ at $r_c=0.5a$.  }
\label{fig:qprofiles}
\end{figure}

The toroidicity induced Alfv\'en eigenmode driven by energetic particles is simulated using the parameters defined by the ITPA-EP (International Tokamak Physics Activity-Energetic Particle) group \cite{konies2018benchmark}. 
For this case, the major radius $R_0=10 \; {\rm m}$, minor radius $a=1$ m, on-axis magnetic field $B_{\rm axis}=3$ T, and the ad-hoc equilibrium has been adopted for which the safety factor profile $q(r)=\hat{q}/\sqrt{1-r^2/R_0^2}$ so that the poloidal flux function $\psi$ can be analytically obtained. There are three choices for the safety factor profiles with an analytical solution to $\psi$ in the TRIMEG code. 
\begin{enumerate}
    \item The second order polynomial $q$ profile in the ITPA-TAE case \cite{konies2018benchmark}  
    \begin{eqnarray}\label{eq:qTAE}
        \bar{q}(r)=\bar{q}_0+\bar{q}_2r^2 \;\;, \;\;
        \psi=\frac{B_0}{2 \bar q_2 }\ln\left(1+\frac{\bar{q}_2}{\bar{q}_0} \rho^2 \right)
    \end{eqnarray}
    where $\bar{q}_0=1.71$, $\bar{q}_2=0.16$. This ad hoc equilibrium has also been implemented in the ORB5 code \cite{lanti2020orb5}.  The corresponding TAE for the benchmark is characterized by the toroidal and poloidal mode numbers $n=-6$ and $m=10, 11$, respectively. 
    \item The second order polynomial $q$ profile in Eq.~\ref{eq:qTAE} for the beta-induced Alfv\'en Eigenmode (BAE), for which $nq=-m$ at $r=0.5a$. In this work, we choose $n=-6$, $\bar{q}_0=1.96, \;1.94, \;1.92,\; 1.90$, $\bar{q}_2=0.16, \;0.24,\; 0.32,\; 0.40$.
    \item The fourth order polynomial $q$ for Reversed Shear Alfv\'en Eigenmode (RSAE) 
    \begin{eqnarray}
        \bar{q}(r)=\bar{q}_0+\bar{q}_2r^2+\bar{q}_4r^4 \;\;,
    \end{eqnarray}
    \begin{eqnarray}
        \psi=\frac{B_0}{\sqrt{4\bar{q}_0\bar{q}_4-\bar{q}_2^2}}\arctan\left( \frac{\bar{q}_2+2\bar{q}_4\rho^2}{\sqrt{4\bar{q}_0\bar{q}_4-\bar{q}_2^2}} \right)
      - {\frac{B_0}{\sqrt{4\bar{q}_0\bar{q}_4-\bar{q}_2^2}}}\arctan\left( \frac{\bar{q}_2}{{\sqrt{4\bar{q}_0\bar{q}_4-\bar{q}_2^2}}} \right)\;\;,
    \end{eqnarray}
    where $\bar{q}_0=1.765625,\; 1.790625,\; 1.790625,\; 1.790625 ,\;1.815625$, $\bar{q}_2=-0.125$ and $\bar{q}_4=0.25$. This analytical ad hoc equilibrium has been implemented in the TRIMEG code for the RSAE studies \cite{meng2022mode} related to the non-perturbative EP effect on symmetry breaking \cite{meng2020effects}. 
\end{enumerate}
The $q$ profiles used in the work are shown in Fig.~\ref{fig:qprofiles}. The $q$ profiles shown in the left frame are for the simulations of TAE and BAE. The $q$ profiles shown in the right frame are for RSAE simulations. 

%%%%%%%%%%%%%%%%%%%%%%%%%%%%%%%%%%%%%%%%%%%%%%%%%%%%%%%%%%%%%%%%%%%%%%

The electron density and temperature are taken to be uniform, with $n_{{\rm e}0}=2.0\times10^{19}\;\rm{m}^{-3}$, $T_{\rm{e}}=1$~keV. The ratio of the electron pressure to the magnetic pressure is $\beta_{\rm e}\approx 9\times 10^{-4}$.  The ion temperature is also uniform, with $T_{i}=1$~keV, and a hydrogen plasma is assumed.  The thermal velocity is defined as $v_{th,s}=\sqrt{2T_s/m_s}$.  The corresponding thermal ion Larmor radius is $\rho_{\rm{ti}}=m_{\rm{i}}v_{{ th},i}/(eB_{\rm{axis}})=1.52\times10^{-3}$~m. To reduce computational cost, a reduced mass ratio $m_{\rm i}/m_{\rm e}=183.6$ is adopted. Fast particles are modeled as deuterons with $m_f/m_i=2$. No poloidal Fourier filter is used in this work.  
 
 The EP density profile is given by 
\begin{eqnarray}
\label{eq:nEP1d}
	n_{\rm{EP}}(r)&=&n_{\rm{EP},0}c_3 \exp\left[ -\frac{c_2}{c_1} \tanh\left(\frac{{r}-c_0}{c_2}\right)\right]\;\;, \\
	\frac{{\rm d}\ln n_{\rm{EP}}}{{\rm d} r} &=&\cosh^{-2}\left(\frac{ r-c_0}{c_2}\right)\;\;,
\end{eqnarray}
where the normalized radial-like coordinate $r=\sqrt{(\psi-\psi_{\rm axis})/(\psi_{\rm edge}-\psi_{\rm axis})}$, $ n_{\rm{EP},0} =1.44131\times10^{17}\; \rm{m}^{-3}$, the subscript `$\rm{EP/f}$' denotes energetic particles (EPs), also referred to as fast ions. The coefficients are $c_0 = 0.491 23$, $c_1 =0.298 228$, $c_2 =0.198 739$, and $c_3 =0.521 298$. This EP density profile is used as the reference case, denoted by $n_{f,ref}$,  where ``ref” indicates the reference case in the following studies. At the radial location $r=c_0$, the EP density divided by the electron density is $n_{EP}(r=0.5a)/n_e=0.0037567601019$. The electron density is radially uniform, and the thermal ion density is given by $n_i(r)=n_e(r)-n_{\rm EP}(r)$ to ensure charge neutrality. In the base case, the EP temperature is $400\; {\rm keV}$. The $n=-6$ mode is simulated by applying a toroidal Fourier filter. An initial perturbation in the marker weights is applied with two poloidal harmonics with $m=10$ and $11$. A total of 
 $32\times10^6$, $8\times10^6$, and $[8\sim16]\times10^6$  markers are used for electrons, ions, and energetic particles, respectively. The radial and poloidal grid numbers are $(N_r,N_\theta)=(64,128)$. For single $n$ simulations without the zonal component, there is only 1 degree of freedom in the toroidal direction by using the particle-in-Fourier method. For cases with the zonal component, the degree of freedom in the toroidal direction is $2$. 
The growth rates and frequencies obtained for different values of $T_{\rm EP}$ benchmarked against other codes, including ORB5, CKA-EUTERPE, and GYGLES in our previous work, showing good agreement \cite{lu2025trimeg}.

Two sets of cases are defined in Tab.~\ref{tab:caseAB}. For Case A, the thermal ions, electrons, and EPs are treated nonlinearly. In Case B, only the EPs are treated nonlinearly, while the thermal ions and electrons are pushed along their unperturbed trajectories. The corresponding solved equations are explained in the caption of Tab.~\ref{tab:caseAB}. 

\begin{table}
\caption{The simulation cases with different choices of linear/nonlinear thermal ions/electrons and EPs. In the nonlinear treatment, Eqs.~\ref{eq:dR0dt}--\ref{eq:du1dt} and Eq.~\ref{eq:dfdt} are solved. In the linear treatment, Eq.~\ref{eq:dR0dt}, Eq.~\ref{eq:du0dt} are solved, Eq.~\ref{eq:dR1dt}, Eq.~\ref{eq:du1dt} are set to zero, and $\delta\dot{\boldsymbol R}$ and $\delta\dot{u}_\|$ are set to zero on the left hand side of Eq.~\ref{eq:dfdt} (the right hand side of Eq.~\ref{eq:dfdt} remains the same). For the linear treatment, the pullback procedure is not applied to the parallel velocity because the particles are pushed along unperturbed trajectories; thus, Eq.~\ref{eq:pullback_v} is omitted. 
}
\centering
\begin{tabular}{l c c c}
\hline
Case & thermal ion & electron & EPs  \\
\hline
 A & NL & NL & NL  \\
\hline
 B & L  &  L & NL  \\
\hline\hline
\end{tabular} \label{tab:caseAB}
\end{table}

\section{The linear and nonlinear non-perturbative effects of EPs and thermal plasmas}
\label{sec:mode_property}

\begin{figure}
 \centering
        \includegraphics[width=0.96\textwidth]{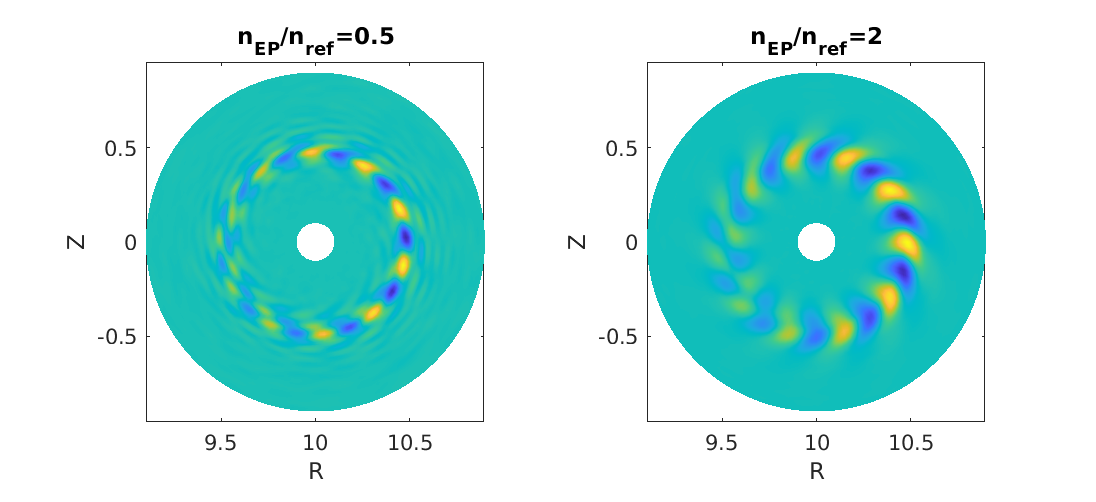}
 \caption{The 2D mode structures of the perturbed density for $n_{EP}/n_{\rm ref}=0.5$ (left) and  $n_{EP}/n_{\rm ref}=2$ (right).}
\label{fig:mode2dnEP}
\end{figure}

Early studies of EP-driven instability assume that the AE mode structure and the frequency are determined by the MHD model, and EPs introduce perturbative corrections to the growth rate and the frequency while the mode structure remains unchanged. While these perturbative models have provided important results on mode growth rates and saturation levels in the studies of tokamak plasmas \cite{pinches1998hagis} and stellarators \cite{slaby2018effects}, the non-perturbative effects of EPs and thermal ions/electrons can be important. The non-perturbative effects of EPs have been discussed in later work \cite{ma2015global,lu2018kinetic}.
The non-perturbative effects of EPs on the mode structure have been observed in gyrokinetic simulations \cite{wang2013radial}. 
We first study the impact of the EP drive on the mode width. As shown in Fig.~\ref{fig:mode2dnEP}, the radial width of the mode structure increases as the EP density increases. The left and middle frames show the 2D structures of the density perturbation for  $n_{EP}/n_{\rm f, ref}=0.5$ (left) and  $n_{EP}/n_{\rm f, ref}=2$ (right). 
Regardless of whether the thermal nonlinearity is included, the change in the radial width is minor, suggesting that the EPs are the key ingredient for the radial broadening of the mode structure in the right frame for this ITPA-TAE case. This EP effect on the radial width can be understood by considering the theoretical studies of the kinetic effects of EP-driven instabilities \cite{chen2016physics,zonca2014theory}.

%%%%%%%%%%%%%%%%%%%%%%%%%%%%%%%%%%%%%%%%%%%%%%%%%%%%%%%%%%%%%%%%%%%%%%

\begin{figure*}
 \centering
    \includegraphics[width=0.45\textwidth]{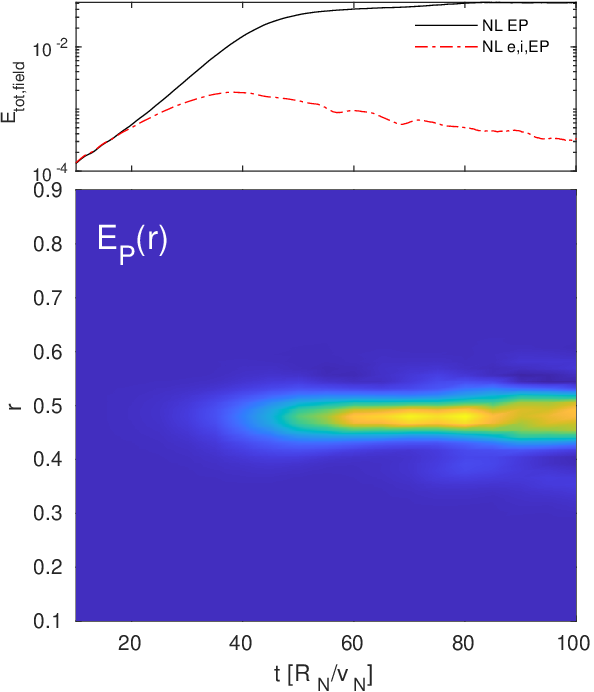}
    \includegraphics[width=0.45\textwidth]{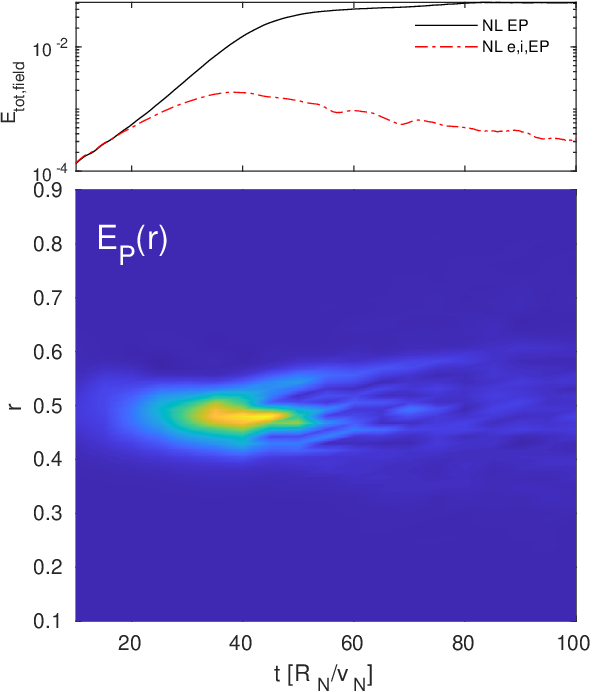}
    \caption{The time evolution of the total field energy integrated over the whole volume (upper) and the time evolution of the radial profile of the electrostatic energy $E_P(r)$ (lower). Lower left: thermal ions and electrons are treated linearly, but EPs nonlinearly (Case B). Lower right: all species are treated nonlinearly (Case A). }
\label{fig:Ert2d}
\end{figure*}

The non-perturbative nonlinear effect of the thermal plasma (electrons and ions) on the time evolution of the total field energy and the mode structure is studied as shown in Fig.~\ref{fig:Ert2d}. The total field energy is calculated as follows,
\begin{eqnarray}
    E_{\rm tot,field}&=&E_P+E_A \;\;, \\
    E_P&=& -\int_\Omega dV \delta \Phi\delta n_v \;\;,\\
    E_A&=& -\int_\Omega dV \delta A_\|\delta j_{\|,v} \;\;,
\end{eqnarray}
where $\delta n_v$ and $\delta j_{\|,v}$ are calculated using the perturbed distribution in $({\bf R},v_\|,\mu)$ coordinates, and $\Omega$ is the total volume or the volume of the local anulus depending on whether the field energy is calculated in the whole volume or locally for the radial profile. 
Case A and Case B defined in Tab.~\ref{tab:caseAB} are simulated using the base parameters ($T_{\rm EP}=400\; {\rm keV}$). For Case A, thermal ions, electrons, and EPs are treated fully nonlinearly, while for Case B, thermal ions and electrons are treated linearly and EPs are treated nonlinearly (the solved equations are described in the caption of Tab.~\ref{tab:caseAB}). In the upper frame of Fig.~\ref{fig:Ert2d}, the time evolution of the linear stage ($t<20 R_N/v_N$) is identical for Case A and Case B. When thermal plasma nonlinearity is absent, the nonlinear stage exhibits a sub-exponential upward drift in which the field energy increases gradually, which has also been observed in the simulations using EUTERPE and FLU-EUTERPE without the thermal plasma nonlinearity \cite{cole2017toroidal}. When the thermal plasma nonlinearity is included, the total field energy reaches a maximum level (``overshoot'') at the end of the nonlinear stage and decreases in the later stage, as shown in the right upper frame. The field energy along the radial direction $E_P(r)$ is shown in the lower frame. For Case B, the mode structure remains the same from the linear to the nonlinear stage, and the radial structure of the field energy  $E_P(r)$ remains peaked near $r=0.5$. For Case A, the nonlinear effect of the thermal ions and electrons broadens the radial width of $E_P(r)$. 

%%%%%%%%%%%%%%%%%%%%%%%%%%%%%%%%%%%%%%%%%%%%%%%%%%%%%%%%%%%%%%%%%%%%%%
%%%%%%%%%%%%%%%%%%%%%%%%%%%%%%%%%%%%%%%%%%%%%%%%%%%%%%%%%%%%%%%%%%%%%%

\begin{figure*}
 \centering
    \includegraphics[width=0.75\textwidth]{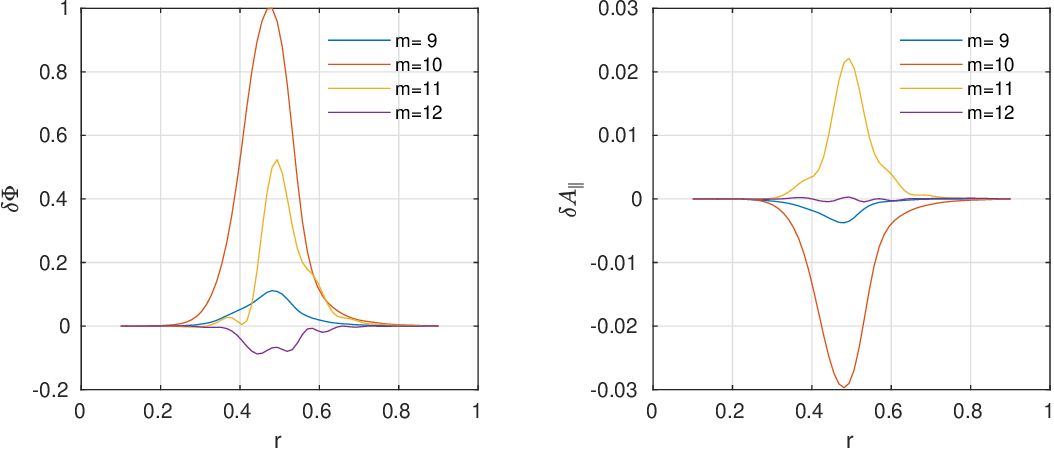}
    \includegraphics[width=0.75\textwidth]{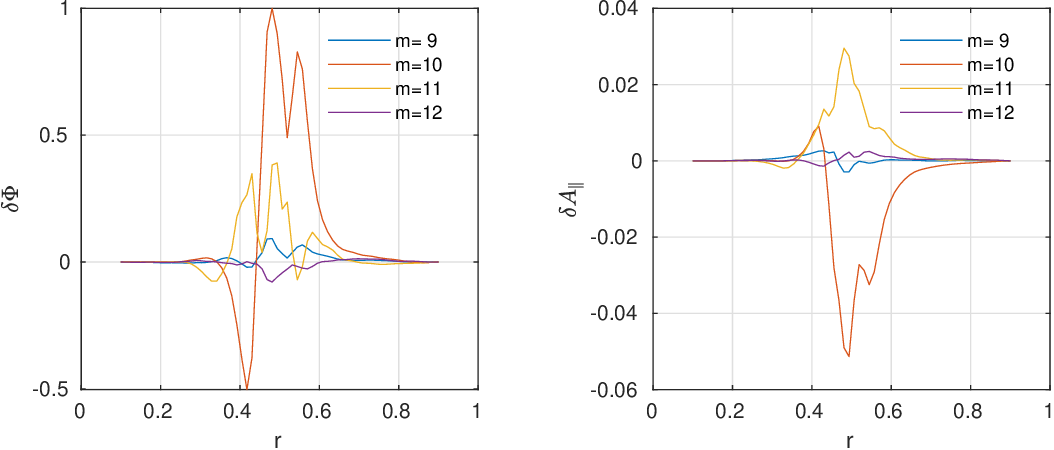}
    \caption{Upper: the radial structures of the poloidal harmonics $\delta\Phi_m(r)$ (left) and $\delta{A}_{\parallel,m}(r)$ (right).
    Lower: the radial structures in the nonlinear stage at $t=50 R_N/v_N$ for Case A (nonlinear electrons, ions, and EPs). }
\label{fig:fmr1d}
\end{figure*}

The radial structure of the poloidal harmonics in the linear stage is shown in the upper frame of Fig.~\ref{fig:fmr1d}. The $m=10, 11$ harmonics are the dominant components. The $\delta\Phi_{10}$ and $\delta\Phi_{11}$ have the same sign, which is referred to as the ``even'' parity. The  $\delta{A}_{\|10}$ and $\delta{A}_{\|11}$ have the opposite sign, which is referred to as the ``odd'' parity. The upper frame describes the typical mode structure of Case A in the linear stage and Case B in both the linear stage and the nonlinear stage.
For Case A, the dominant harmonics split at the peak of the radial structure at the stage near the nonlinear saturation, as shown in the lower frame. Apparently, the  nonlinear effect of the thermal ion and electron changes the mode structure significantly during the nonlinear stage. 

%%%%%%%%%%%%%%%%%%%%%%%%%%%%%%%%%%%%%%%%%%%%%%%%%%%%%%%%%%%%%%%%%%%%%%
%%%%%%%%%%%%%%%%%%%%%%%%%%%%%%%%%%%%%%%%%%%%%%%%%%%%%%%%%%%%%%%%%%%%%%

\section{Thermal plasma effects on EP density flattening and particle flux}
\label{sec:transport}

\begin{figure}
 \centering
        \includegraphics[width=0.85\textwidth]{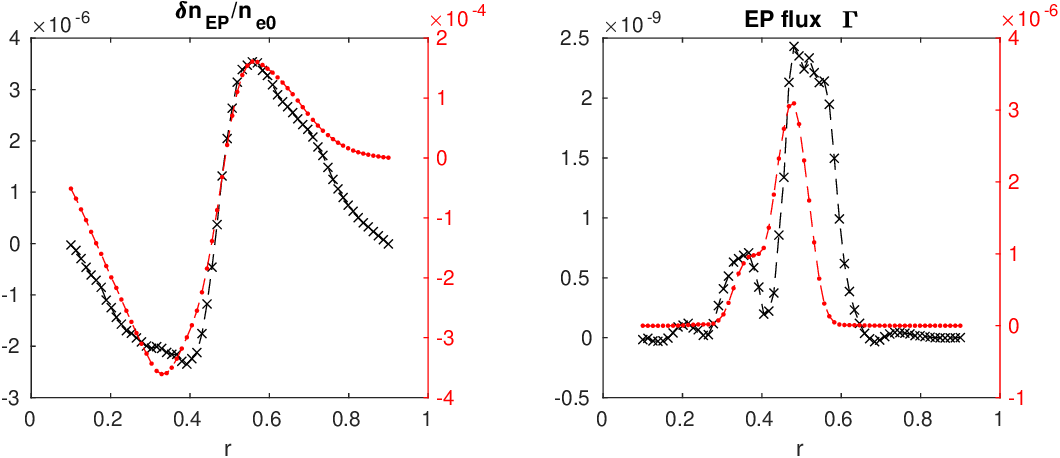}
 \caption{The change of the EP density (left) and the EP particle flux (right) for the simulations with only EP nonlinearity and the nonlinearities of all species.}
\label{fig:EPflux_nf}
\end{figure}

The effects of thermal ions and electrons on the particle flux of energetic particles are studied as shown in Fig.~\ref{fig:EPflux_nf}.
Case A and Case B have been compared for the base parameter $n_f/n_{ref}=1$.
%In Cases A and B, TAE frequency does not vary. 
The particle flux $\Gamma$ is calculated according to 
\begin{eqnarray}
\label{eq:flux_n}
\Gamma =\left\langle\int {\rm d}^3v \frac{\boldsymbol b}{B}\times\langle\nabla\delta\phi-v_\|\delta{A}_\|\rangle\delta f_v \right\rangle_{F}\;\;, \;\;  
    %\Gamma_{\delta\phi} =\left\langle\int {\rm d}^3v \frac{\boldsymbol b}{B}\times\langle\nabla\delta\phi\rangle\delta f_v \right\rangle_{F}\;\;, \;\;    \Gamma_{\delta{A}_\parallel}    =\left\langle\int {\rm d}^3v \frac{\boldsymbol b}{B}\times\langle\nabla\delta{A}_\parallel\rangle \delta f_v  v_\|\right\rangle_{F} \;\;,
\end{eqnarray}
where $\delta f_v$ is the perturbed distribution in $({\boldsymbol R},v_\|,\mu)$ coordinates, $\langle\cdots\rangle$ denotes the gyro-average when the finite Larmor radius effect is included consistently, and $\langle\ldots\rangle_F$ denotes the flux surface average. 
The particle flux is normalized to the gyro-Bohm scaling with $\Gamma_{\rm GB}=(\rho_{\rm ref}/a)^2v_Nn_{e}$, where  $v_N=\sqrt{2T_N/m_N}$, $T_N=T_e=1$ keV, and $m_N$ is the proton mass. 

The fluxes in the nonlinear stage are shown in Fig.~\ref{fig:EPflux_nf}. As shown in the left panel of Fig.~\ref{fig:EPflux_nf}, the flattening of the EP density profile occurs in both cases, and the flattened regions are relatively wide. However, the magnitude of the density variation is much smaller in Case A than in Case B, which is consistent with the comparison of the saturation levels.
Notably, the locations where $\delta n_{EP}=0$ are slightly shifted between the two cases, and the $\delta n_{EP}$ profile is not symmetric around these points, which is related to the particle loss at the outer boundary at the higher saturation level of the instability.

In the right panel, the EP flux remains positive across the radial domain, indicating that the EP transport is predominantly outward and consistent with the relaxation of the EP density gradient. A clear difference in flux magnitude is observed between the two cases, with the flux in Case A being significantly smaller than in Case B. For Case A (black line), two dominant peaks are clearly visible, indicating the presence of two resonance locations, consistent with the splitting of the mode structure during the nonlinear stage. In addition, the flux region is broader than in Case B.

Overall, despite its lower flux magnitude, Case A exhibits a more structured, multi-peak profile compared with Case B, suggesting that EP transport is distributed over multiple radial locations rather than concentrated in a single dominant region.
%%%%%%%%%%%%%%%%%%%%%%%%%%%%%%%%%%%%%%%%%%%%%%%%%%%%%%%%%%%%%%%%%%%%%%
%%%%%%%%%%%%%%%%%%%%%%%%%%%%%%%%%%%%%%%%%%%%%%%%%%%%%%%%%%%%%%%%%%%%%%

\begin{figure}
 \centering
        \includegraphics[width=0.85\textwidth]{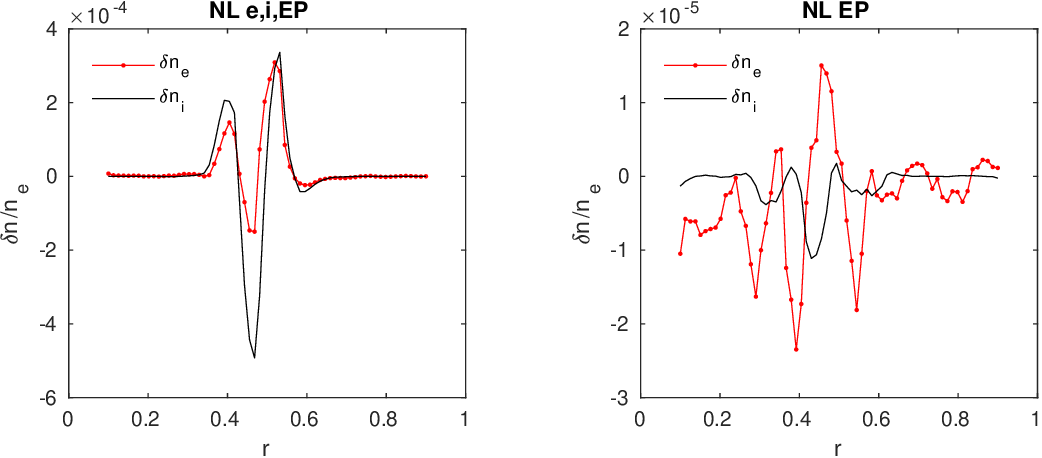}
 \caption{The flux-surface averaged density perturbation with (left) and without (right) thermal plasma nonlinearity.}
\label{fig:deltanei}
\end{figure}

As shown in Fig.~\ref{fig:deltanei}, the density perturbation profiles of thermal ions and electrons in Case A and Case B are significantly different. Although Case B reaches a much higher saturation level, the density perturbations of thermal ions and electrons are smaller, since their response mainly reflects the weight change along the unperturbed particle orbits.

In contrast, in Case A, the ions and electrons move along perturbed orbits, which leads to larger density perturbations.
In Case A, $\delta n_i$ and $\delta n_e$ are larger than the EP density perturbation, indicating a strong effect of the TAE mode on the thermal particles' redistribution in the radial direction.
In addition, the perturbed ion and electron densities exhibit similar profiles, so their contributions to the net charge largely cancel out, even if the $n=0$ field solver is switched on.
In Case B, however, $\delta n_i$ and $\delta n_e$ are smaller than the EP density variation associated with the EP profile flattening. Therefore, the mode structure is mainly determined by the EP redistribution with minor contributions from thermal ions and electrons.

%In Case A, the collective nonlinear response of thermal ions and electrons suggests that the particles adjust their motion to maintain charge quasineutrality. In Case B, the $\delta n_i$ and $\delta n_e$ are small.
%%%%%%%%%%%%%%%%%%%%%%%%%%%%%%%%%%%%%%%%%%%%%%%%%%%%%%%%%%%%%%%%%%%%%%
%%%%%%%%%%%%%%%%%%%%%%%%%%%%%%%%%%%%%%%%%%%%%%%%%%%%%%%%%%%%%%%%%%%%%%

\section{Phase space analyses}
\label{sec:phase_space}
\begin{figure}
 \centering
        \includegraphics[width=0.98\textwidth]{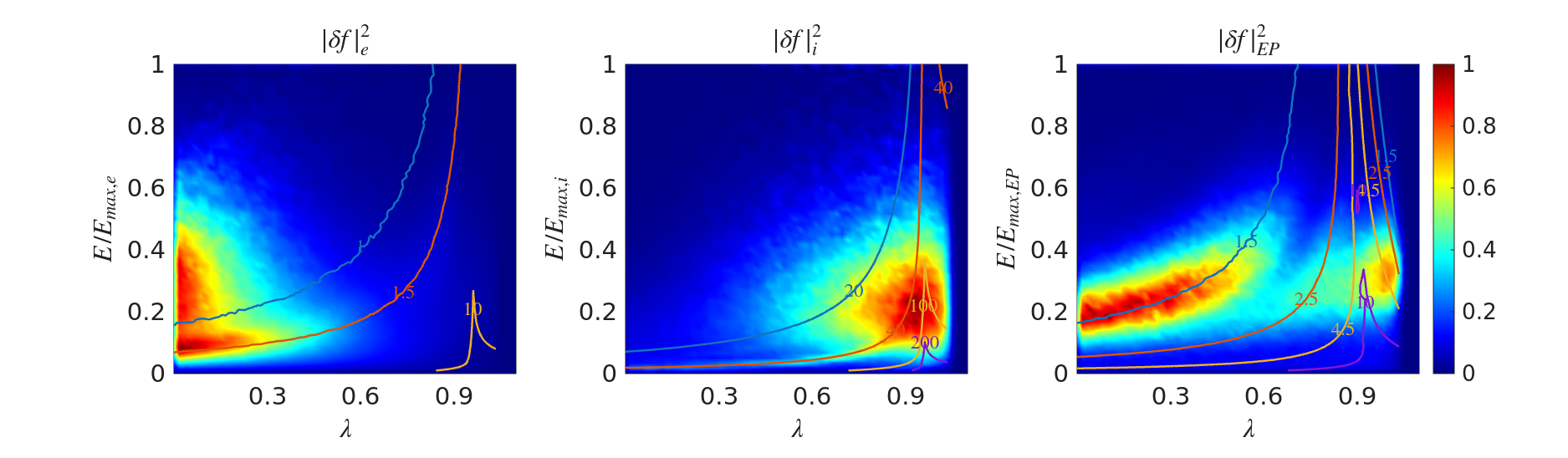}
        \includegraphics[width=0.98\textwidth]{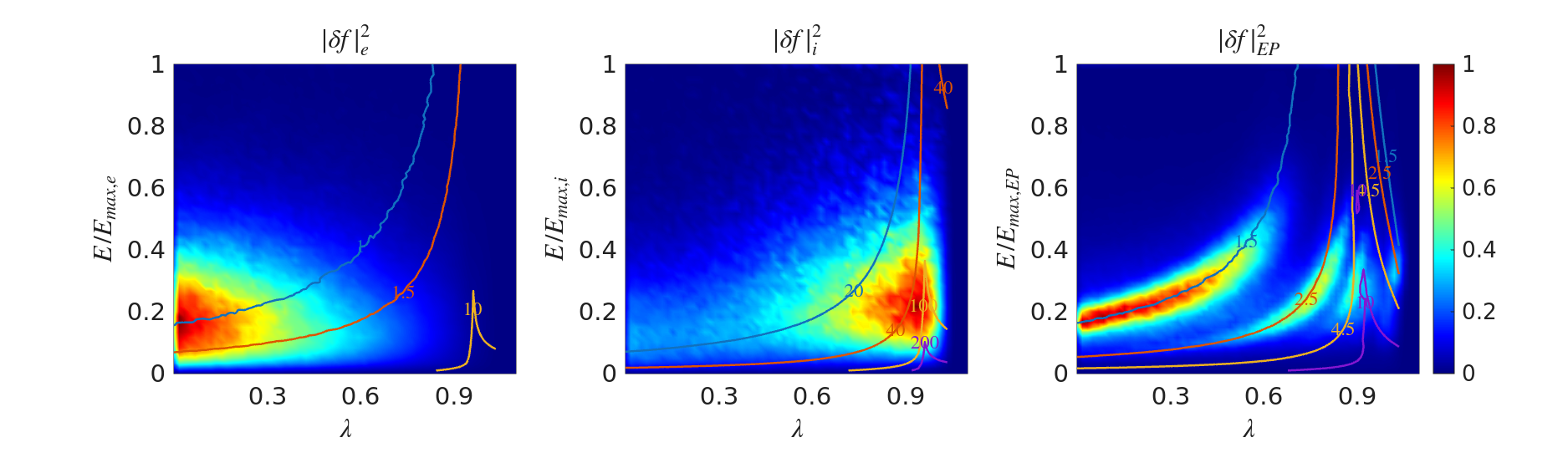}
        \includegraphics[width=0.98\textwidth]{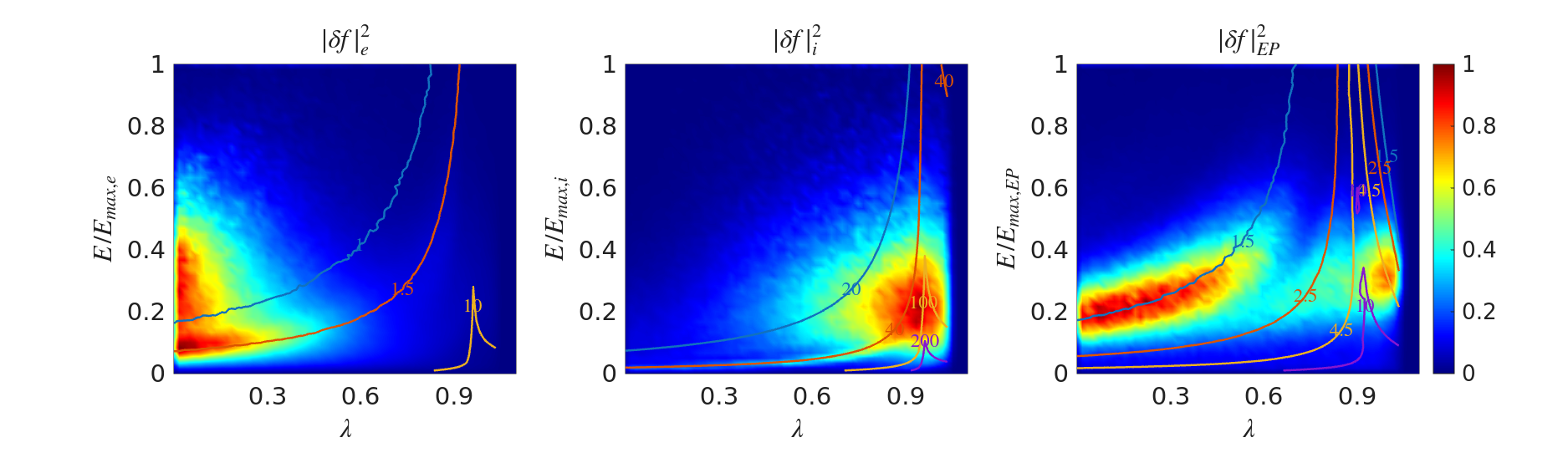}
        \includegraphics[width=0.98\textwidth]{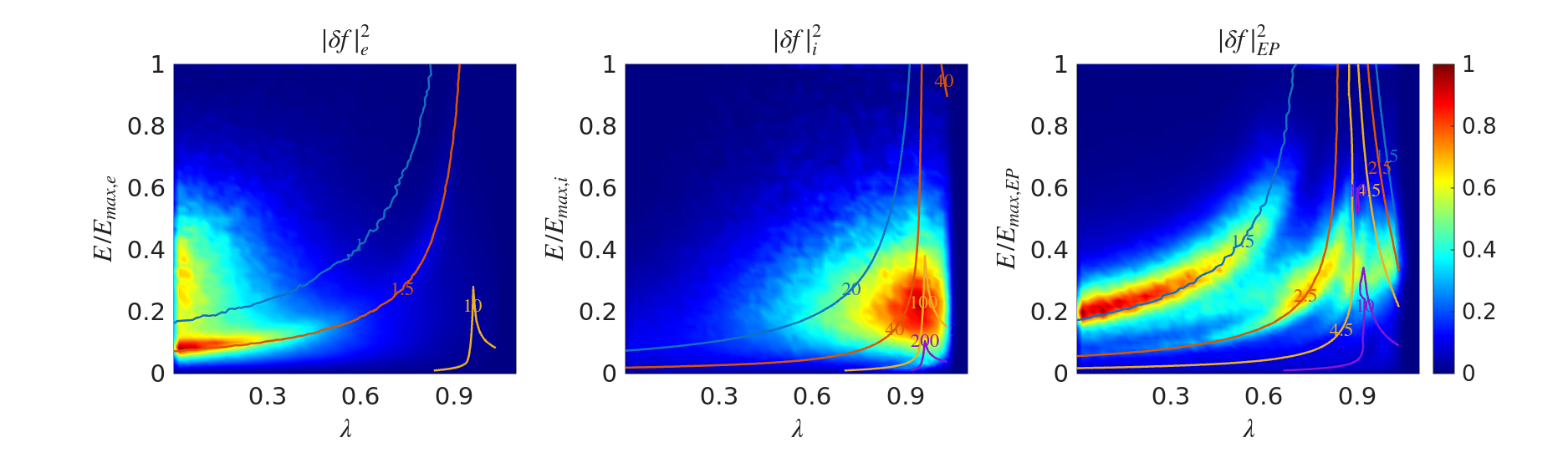}
 \caption{The resonance structure of the square of the perturbed distribution function in $(E,\lambda)$ space. The results in the linear stage (first row) and nonlinear stage (second row) with thermal plasma nonlinearity (Case A). The results in the linear (third row) and the nonlinear (fourth row) stage without thermal plasma nonlinearity (Case B). The bounce-precession resonance condition in Eq.\eqref{eq:omega_res_precession} is shown for $p$=1, 1.5, 10 (electrons), $p=$20, 40, 100, 200 (ions), and $p=$1.5, 2.5, 4.5, 10 (EPs).}
\label{fig:resonance}
\end{figure}

The phase-space structure of the perturbed distribution function is analyzed to identify the wave-particle interactions using the previous implementation of the diagnostic tools in the TRIMEG code \cite{li2026gyrokinetic}. The square of the perturbed distribution function is shown in Fig.~\ref{fig:resonance}, where $E/E_{max,s}$ is the normalized particle energy (with $E_{max,s}=v_{max,s}^2/2$ the species-dependent maximum energy, $v_{max,e}=3v_{th,e}$, $v_{max,i}=3v_{th,i}$, $v_{max,EP}=2v_{th,EP}$), and $\lambda=\mu B_{\rm axis}/E$ characterizes the pitch angle. The phase structure in the linear regime is the same regardless of the inclusion of the thermal nonlinearity, as indicated by the first row (linear stage, with thermal nonlinearity) and the third row (linear stage, without thermal nonlinearity). 
The linear resonance condition is given by 
\begin{equation}\label{eq:omega_res}
    n\omega_{\zeta}+l\omega_{b}-\omega=0,
\end{equation}
where $n$ is the toroidal mode number and  $l$ is an integer. For this TAE, $n=-6$ and the mode frequency is $\omega \approx 4\times10^5 \; {\rm rad/s}$.
The linear resonance condition only depends on the equilibrium. The transit (for passing particles) or bounce (for trapped particles) frequency is defined as
\begin{equation}
    \omega_b=\frac{2\pi}{\oint \mathrm{d}\theta/\dot{\theta}} \;.
\end{equation}
The toroidal transit frequency is given by
\begin{equation}
\omega_{\zeta}= \frac{\omega_b}{2\pi}\oint \dot{\zeta}\frac{\mathrm{d}\theta}{\dot{\theta}},
\end{equation}
which represents the time-averaged change in the toroidal angle $\Delta\zeta$ over one poloidal transit (or bounce) period \cite{meng2018resonance}. In general, the integer $l$ is close to the poloidal mode number $m$ of the perturbation, but the exact resonance location can be found only by integrating over the particle poloidal trajectory \cite{white2013}. The precession (drift) frequency is defined by subtracting the field-line-following motion, i.e.,
\begin{equation}
    \omega_{d}= \frac{\omega_b}{2\pi}\oint( \dot{\zeta} - \bar{q} \dot{\theta} )\frac{\mathrm{d}\theta}{\dot{\theta}},
\end{equation}
where $\bar{q}$ is time-averaged safety factor.
%if r, zeta, theta is straght field line coordinates, along field-line, $q=d\zeta/d\theta$.
For passing particles, this reduces to $\omega_{d}=\omega_\zeta-\bar{q}\omega_b$ while for trapped particles $\omega_{d}=\omega_\zeta$. 
The linear bounce-precession resonance condition is given by
\begin{equation}\label{eq:omega_res_precession}
    n\omega_{d}+p\omega_{b}-\omega=0.
\end{equation}
Since the TAE features two dominant poloidal harmonics $m$, passing particles exhibit the well-known half-integer $p$ resonances \cite{chen2016physics}. In Fig.~\ref{fig:resonance}, we plot the linear resonance lines given by Eq.~\eqref{eq:omega_res_precession}, as they show better agreement with the dominant resonance structure of $|\delta f|^2$ from simulations \cite{li2026gyrokinetic}. This clearly reveals the underlying bounce-precession resonance of passing EPs with the TAE. The integer $p$ is shown on each resonance line. 

In the linear stage, the distributions of $|\delta f|^2$ are nearly identical for Case A (first row) and Case B (third row).  For EP, the dominant resonance in phase space is aligned with $p=3/2$, while resonances in the trapped region consist of multiple overlapping resonances with different $p$.
Notably, electrons also exhibit a $p=3/2$ resonance. In contrast, ions have much smaller characteristic frequencies than the wave, limiting resonance to those near the passing-trapped boundary.

In Case A, during the nonlinear stage (second row), the EP-wave resonance becomes more pronounced, and the $p=5/2$ emerges. 
The values $p=1.5,\; 2.5$ are consistent with the observations in the previous benchmark work \cite{konies2018benchmark}, which show that the parallel velocity of resonant EPs is near $v_{\rm A}/3$ and $v_{\rm A}/5$, respectively, where $v_{\rm A}$ is the Alfvén velocity. Since $\omega_d\ll \omega$, the resonance condition reduces to approximately $p\omega_b=\omega$ for the resonance of passing particles.  For well-passing particles, $\omega_b\approx v_\parallel/(qR)$, while the TAE frequency satisfies $\omega =k_\parallel v_{\rm A}$ with $k_\parallel\approx 1/(2qR)$. These relations yield $v_\parallel\approx v_{\rm A}/3$ for $p=1.5$, and $v_\parallel=v_{\rm A}/5$ for $p=2.5$. 
A similar behavior is observed in Case B (fourth row). However, in Case B,  the EP resonances exhibit stronger overlap, indicating that more particles participate in the interaction over a broader phase-space region. This resonance overlap is likely associated with the larger mode amplitude.
In the nonlinear saturation stage, the electron phase-space structures differ significantly between Case A (second row) and Case B (fourth row). In Case A, the electron $|\delta f|^2$ distribution is broader in both $E$ and $\lambda$, whereas in Case B it is primarily concentrated around the $p=3/2$ resonance. 
This suggests that nonlinear electron dynamics play an important role in determining the saturation amplitude of the instability and the nonlinear resonance.

In this context, it is worth emphasizing that phase-space zonal structures (PSZS) play a fundamental role in the underlying transport processes, as indicated in Fig.~\ref{fig:deltanei} in this work and discussed in previous works \cite{zonca2015nonlinear,falessi2019transport,falessi2023nonlinear}. In particular, the concept of a “zonal state” was introduced in \cite{falessi2019transport}, providing a unified framework to describe the self-consistent interaction between fluctuations and zonal structures. Collisional effects can be considered consistently with the evolution of the PSZS to predict EP transport \cite{meng2024NF,lauber2024atep}. Furthermore, PSZS can be regarded as indispensable synthetic diagnostics for the verification and validation of nonlinear gyrokinetic simulations, with the same importance as the underlying fluctuation structures. Therefore, the features observed in Fig.~\ref{fig:resonance} are not only indicative of the fluctuation dynamics but also directly reflect the associated transport processes mediated by PSZS.  The theoretical and numerical analysis of the TAE saturation mechanism with the consideration of the thermal plasma nonlinearity can be found in a dedicated work \cite{chen2026nonlinear}. 

%%%%%%%%%%%%%%%%%%%%%%%%%%%%%%%%%%%%%%%%%%%%%%%%%%%%%%%%%%%%%%%%%%%%%%

\section{The effect of the zonal flow }
\label{sec:zonal}
\begin{figure}
 \centering
        \includegraphics[width=0.85\textwidth]{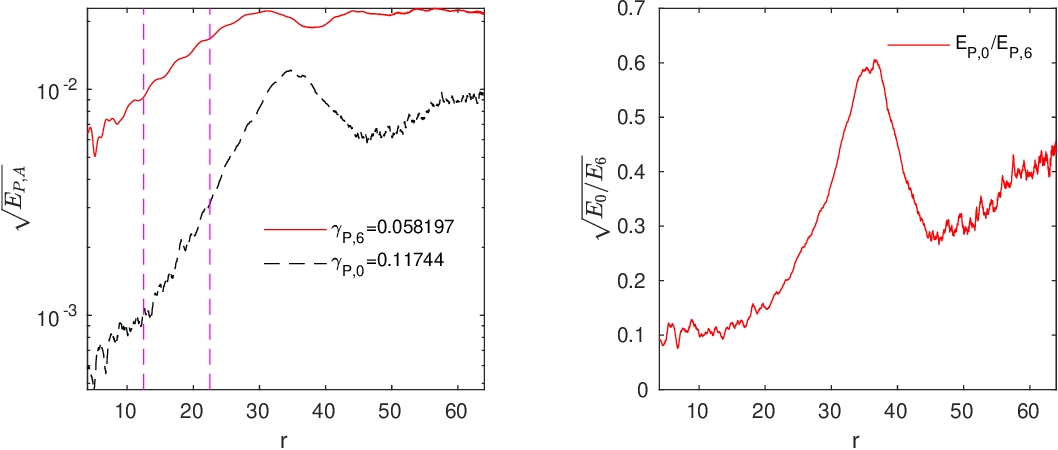}
 \caption{The time evolution of the total field energy in the simulation with $n=0, 6$.}
\label{fig:Etotn0n6}
\end{figure}

Simulations are carried out to evaluate the effect of the zonal components ($n=m=0$) on the saturation and transport level. The $n=-6$ and $n=0$ components are kept in the field solver and the particle pusher, with $(32, 8, 8\sim16)\times10^6$ electrons, thermal ions, and EPs in the simulation. 
The time evolution of the square root of the total field energy is shown in the left frame Fig.~\ref{fig:Etotn0n6}. 
%The total field energy from $\delta\phi_{6}$ and  $\delta{A}_{\|6}$ has a similar magnitude, while that from the zonal component is lower. 
The growth rate is measured using the data between the two dashed vertical lines. The growth rate of the electrostatic part of the zonal component $\gamma_{P,0}=0.11744$ is very close to twice the growth rate of the TAE $2\times0.058197=0.116394$, suggesting the mechanism of the beat-driven zonal flow \cite{chen2024beat}.

The ratio between the zonal component and the TAE is shown in the right frame of Fig.~\ref{fig:Etotn0n6}. Since the zonal component has a larger growth rate, the ratio of zonal field energy to the $n=-6$ field energy increases during the linear stage. The more pronounced overshot of the $E_{p,0}/E_{p,6}$ indicates that $E_{p,6}$ reaches saturation earlier than $E_{p,0}$. 
% In contrast, the corresponding ratio for the magnetic component remains smaller.

\begin{figure}
 \centering
        \includegraphics[width=0.48\textwidth]{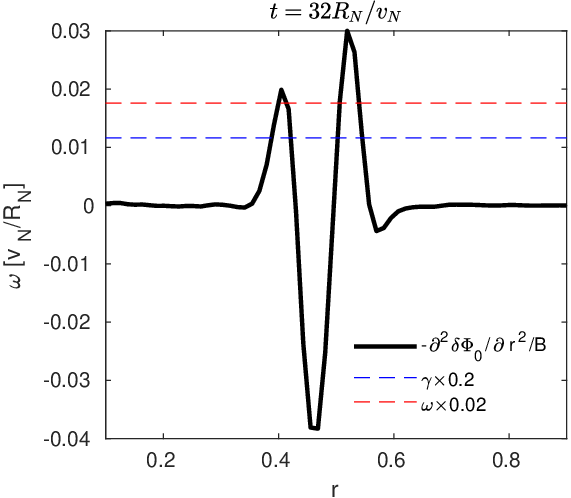}
 \caption{The zonal flow shearing rate $\partial v_{ZF}/\partial r\approx-\partial^2\delta\Phi/\partial r^2/B$ and its comparison with the growth rate and real frequency of the TAE.}
\label{fig:zfshear}
\end{figure}

The strength of the zonal component in the total field energy is moderate ($\sim0.6$) right after the saturation. To further estimate the possible effect of zonal flow on the TAE, we compared the radial gradient of the poloidal zonal flow (the zonal flow shearing rate) as shown in Fig.~\ref{fig:zfshear}. The  zonal flow shearing rate is calculated according to $\partial v_{ZF}/\partial r\approx-\partial^2\delta\Phi/\partial r^2/B$. The zonal flow shearing rate is much smaller than the real frequency and the TAE growth rate.
%, indicating that the zonal flow effect on stabilization is minor. 

%%%%%%%%%%%%%%%%%%%%%%%%%%%%%%%%%%%%%%%%%%%%%%%%%%%%%%%%%%%%%%%%%%%%%%
%%%%%%%%%%%%%%%%%%%%%%%%%%%%%%%%%%%%%%%%%%%%%%%%%%%%%%%%%%%%%%%%%%%%%%

%%%%%%%%%%%%%%%%%%%%%%%%%%%%%%%%%%%%%%%%%%%%%%%%%%%%%%%%%%%%%%%%%%%%%%
%%%%%%%%%%%%%%%%%%%%%%%%%%%%%%%%%%%%%%%%%%%%%%%%%%%%%%%%%%%%%%%%%%%%%%
\section{Stiffness of the saturation and transport level for varying EP drive}
\label{sec:stiffness}
The stiffness of EP transport has been studied in the DIII-D experiment \cite{collins2016observation}, and it is observed that EP transport becomes stiff above a critical threshold in the presence of many overlapping small-amplitude Alfvén eigenmodes. In the following, we refer to ``stiffness of the EP transport'' as the change of the EP flux divided by the change of the EP drive (specifically, the EP density in this work), namely,
\begin{eqnarray}
\label{def:Stiffnessflux}
    S_{\Gamma_{\rm EP}} = \frac{\partial \Gamma_{\rm EP}}{\partial n_{\rm EP}} \;\;.
\end{eqnarray}
The EP profile stiffness, however, can be defined in a different way, namely
\begin{eqnarray}
\label{def:StiffnessnEP}
    S_{P_{\rm EP}} = \frac{\partial P_{\rm EP}}{\partial Src_{\rm EP}} \;\;,
\end{eqnarray}
which describes how the EP pressure $P_{\rm EP}$ can be enhanced as the EP source $Src_{\rm EP}$ is applied, as is consistent with the context in the previous work \cite{collins2016observation}. In our work, we refer to Eq.~\eqref{def:Stiffnessflux} when discussing the stiffness, which has no conflict with Eq.~\eqref{def:StiffnessnEP} when the context is clarified. 

\begin{figure}
 \centering
        \includegraphics[width=0.8\textwidth]{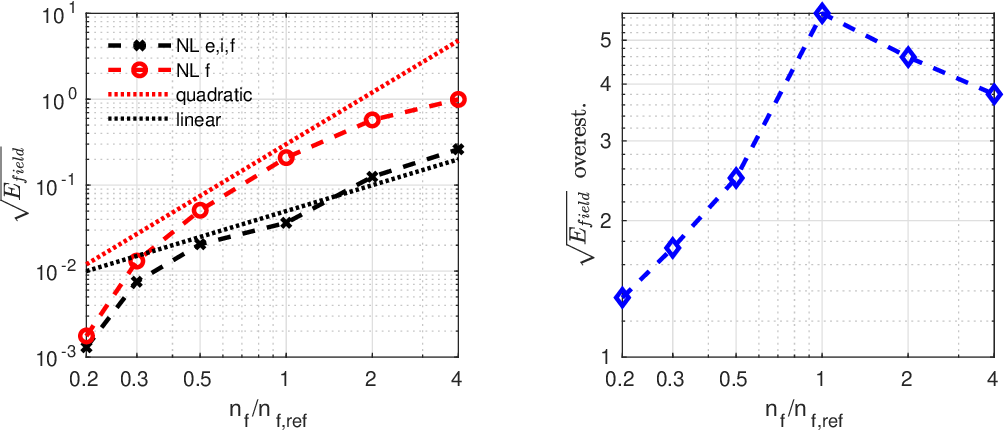}
 \caption{The saturation level of the square root of the total field energy for the simulations with only EP nonlinearity and the nonlinearities of all species.}
\label{fig:saturation_nf}
\end{figure}

The effects of thermal ions and electrons on the saturation of the total field energy are studied as shown in Fig.~\ref{fig:saturation_nf}. 
For cases with only EP nonlinearity (red line), the saturation level can be theoretically explained by the pendulum model, which predicts a quadratic scaling with respect to the linear growth rate. The observation of the quadratic scaling is consistent with the previous studies of resonance broadening width and the validations of the Resonance Broadening Quasilinear theory \cite{gorelenkov2018resonance,meng2018resonance}. 
When the EP drive exceeds a critical value (for this ITPA case, approximately $n_{f,\rm ref}$ as reported in \cite{slaby2018effects}), a transition from quadratic to linear scaling, due to radial decoupling, is also observed with only EP nonlinearity.

However, the saturation level is overestimated.
As the thermal plasma nonlinearity is included (black line), the saturation level is lower compared to the cases with only EP nonlinearity, especially when the EP density is high.

\begin{figure}
 \centering
        \includegraphics[width=0.8\textwidth]{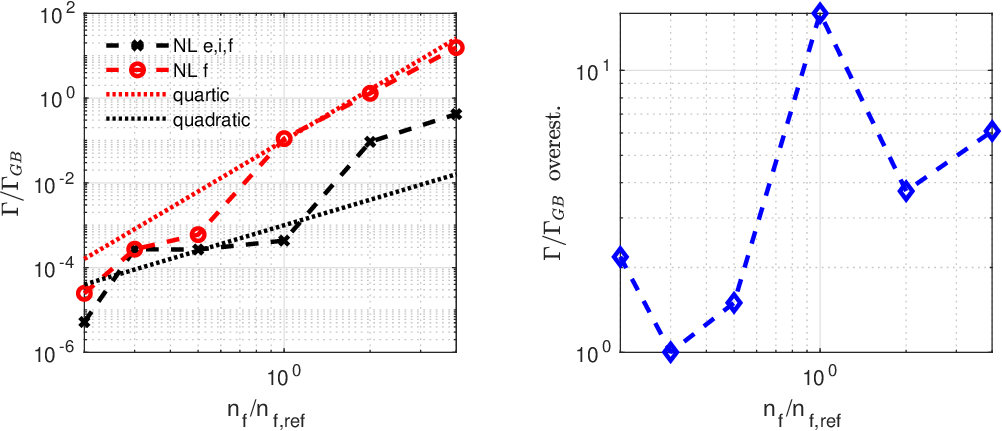}
 \caption{Left: The saturation level of the particle flux of EPs for the simulations with only EP nonlinearity and the nonlinearities of all species. Right: Overestimation factor of the model that excludes thermal plasma nonlinearity.  }
\label{fig:fluxn_nf}
\end{figure}

The EP particle flux for different EP densities is calculated, as shown in Fig.~\ref{fig:fluxn_nf}. 
Since the particle flux is proportional to $|\delta\phi|^2$ and $|\delta{A}_\||^2$, the quartic and quadratic scalings are expected as thermal nonlinearity is excluded or included, respectively. This behavior reflects the quadratic and linear scalings in the saturation level of $\delta \phi$ and $\delta{A}_\|$ as shown in Fig.~\ref{fig:saturation_nf}. As shown in Fig.~\ref{fig:fluxn_nf}, the red solid curve (without thermal nonlinearity) follows the quartic scaling while the black solid curve (with thermal  nonlinearity) follows the quadratic scaling. In short words, when ion and electron nonlinearities are included, the particle flux $S_{\Gamma_{\rm EP}}$ is reduced, indicating less stiffer of EP transport with respect to the EP density,  and a better EP confinement than that predicted by the perturbative model. 

\begin{figure}
 \centering
        \includegraphics[width=0.40\textwidth]{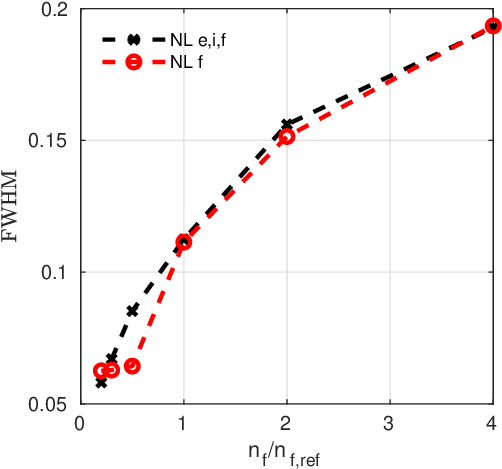}
 \caption{The full width at half maximum (FWHM) of the linear structure of the perturbed density for different values of EP density.}
\label{fig:mode_width}
\end{figure}

The linear scaling of the saturation level due to radial decoupling has been studied in the previous work, where it was shown to depend on the radial width of the Alfv\'enic mode and the finite orbit width of EPs \cite{briguglio2014analysis}. However, the reduced stiffness of the saturation scaling observed in Case A in Fig.~\ref{fig:saturation_nf} is not related to the transition between the resonance detuning and the radial decoupling \cite{briguglio2014analysis}. While thermal plasma nonlinearity plays a significant role in reducing the stiffness of EP transport, their effect on the increase of the radial mode width is relatively small (although finite), as indicated in Fig.~\ref{fig:mode_width}. As discussed in Sec.~\ref{sec:mode_property}, the radial mode width increases as the EP density increases, which is consistent with Fig.~\ref{fig:mode_width}. In addition, the mode width with or without electron and ion nonlinearities are very similar at relatively high EP density ($n_{\rm f}/n_{f,\rm ref}\ge 1$), indicating that the non-perturbative effect on the mode width is primarily due to EPs. 

%%%%%%%%%%%%%%%%%%%%%%%%%%%%%%%%%%%%%%%%%%%%%%%%%%%%%%%%%%%%%%%%%%%%%%
%%%%%%%%%%%%%%%%%%%%%%%%%%%%%%%%%%%%%%%%%%%%%%%%%%%%%%%%%%%%%%%%%%%%%%

%%%%%%%%%%%%%%%%%%%%%%%%%%%%%%%%%%%%%%%%%%%%%%%%%%%%%%%%%%%%%%%%%%%%%%
%%%%%%%%%%%%%%%%%%%%%%%%%%%%%%%%%%%%%%%%%%%%%%%%%%%%%%%%%%%%%%%%%%%%%%

\section{Effects of thermal plasma nonlinearity on the saturation level of BAE}
\label{sec:BAE}
\begin{figure}
 \centering
        \includegraphics[width=0.80\textwidth]{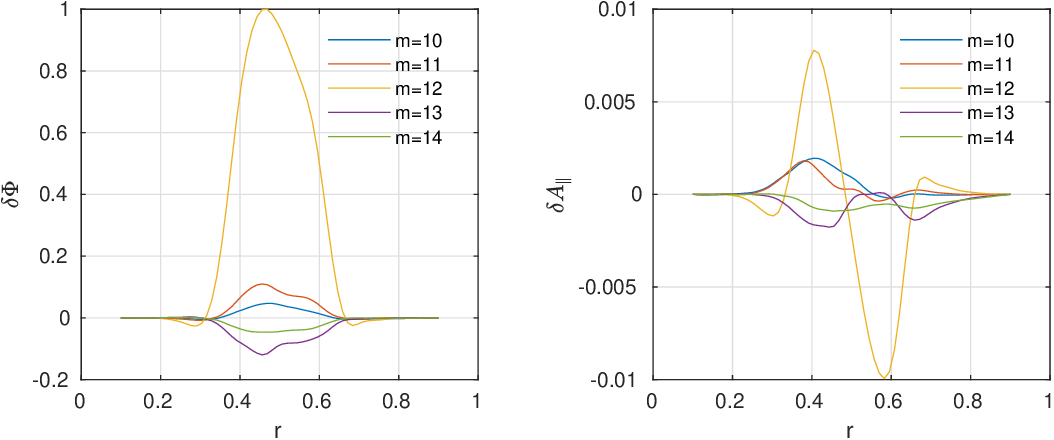}
 \caption{The radial structure of the poloidal harmonics of the Beta induced Alfv\'en Eigenmode (BAE) for $\bar{q}_0=1.96$,  $\bar{q}_2=0.16$.}
\label{fig:fmr_BAE}
\end{figure}

In addition to the TAE studies, the Beta induced Alfv\'en Eigenmode (BAE) is simulated to identify the effects of the thermal plasma nonlinearity on the saturation level. The 2nd type of $q$ profile in Sec.~\ref{sec:models} is adopted with $\bar{q}(r=0.5)=2$ to simulate the $n=-6$ BAE. For the case with $\bar{q}_0=1.96$, the radial structure of the poloidal harmonics is shown in Fig.~\ref{fig:fmr_BAE}. The $\delta\Phi$ and $\delta{A}_\|$ is dominated by the $m=12$ component. $\delta\Phi_{m=12}(r)$ and $\delta{A}_{\|m=12}(r)$ have even and odd parity along the radial direction with respect to $r=0.5$, respectively. 

\begin{figure}
 \centering
        \includegraphics[width=0.4\textwidth]{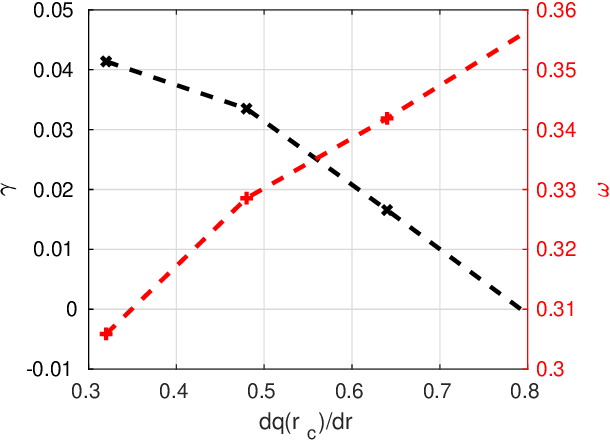}
        \includegraphics[width=0.37\textwidth]{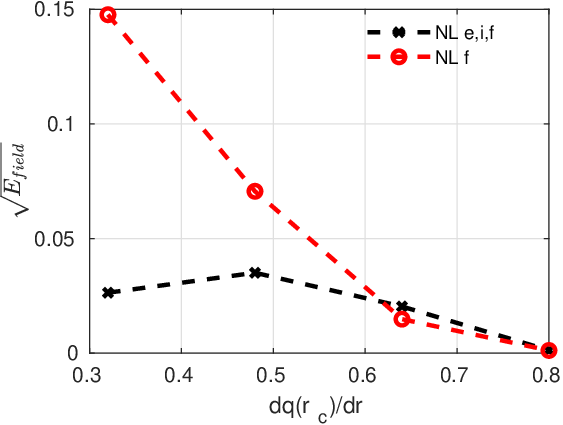}
 \caption{Left: the BAE growth rate and frequency for different  $q$ profiles characterized by the at $r_c$. Right: the saturation level of the BAE with and without thermal plasma nonlinearity. }
\label{fig:saturationBAE}
\end{figure}

The BAE growth rate is sensitive to the $q$ profile, featured by the local magnetic shear at $r_c=0.5a$, as shown in the left frame of Fig.~\ref{fig:saturationBAE}.  The real frequency increases by $\sim20\%$ as $dq(r_c)/dr$ changes from $0.32$ to $0.8$, while the  growth rate decreases with increasing shear $dq(r_c)/dr$. When $dq(r_c)/dr$ reaches $0.8$, the BAE is almost stabilized. The saturation levels are shown in the right frame for cases with and without the thermal plasma nonlinearity. For the low shear case with $dq(r_c)/dr=0.32$, the saturation level is overestimated by $\sim5$ times since the growth  rate and corresponding saturation level are relatively high. For the cases with lower growth rates, the estimate of the saturation level without the thermal plasma nonlinearity is more reliable. This trend is similar to that observed for the TAE case, although the TAE saturation level varies due to the change of the EP density in Sec.~\ref{sec:stiffness}. 

%%%%%%%%%%%%%%%%%%%%%%%%%%%%%%%%%%%%%%%%%%%%%%%%%%%%%%%%%%%%%%%%%%%%%%

%%%%%%%%%%%%%%%%%%%%%%%%%%%%%%%%%%%%%%%%%%%%%%%%%%%%%%%%%%%%%%%%%%%%%%

\section{Effects of thermal plasma nonlinearity on the saturation level of RSAE}
\label{sec:RSAE}
\begin{figure}
 \centering
        \includegraphics[width=0.80\textwidth]{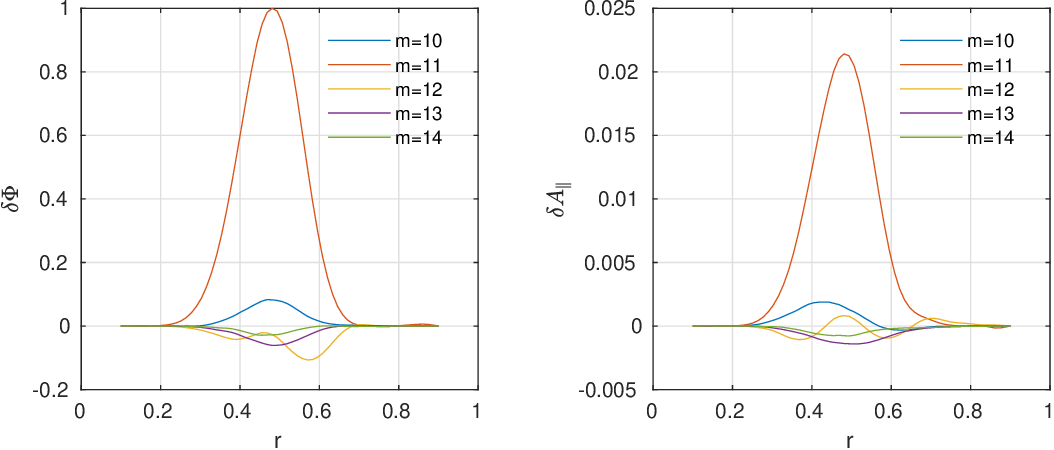}
 \caption{The radial structure of the poloidal harmonics of the Reversed Shear Alfv\'en Eigenmode (RSAE) for $\bar{q}_0=1.815625$,  $\bar{q}_2=-0.125$, $\bar{q}_4=0.25$. Note for $\bar{q}_0=1.765625$, the mode structure is similar to that of the TAE, with the $m=10, 11$ components dominated.  }
\label{fig:fmr_RSAE}
\end{figure}

Following the studies of the BAE, the RSAE is studied. The 3rd type of $q$ profile introduced in Sec.~\ref{sec:models} is adopted to simulate the $n=-6$ RSAE. For the case with $\bar{q}_0=1.815625$, the radial structure of the poloidal harmonics is shown in Fig.~\ref{fig:fmr_RSAE}. The $\delta\Phi$ and $\delta{A}_\|$ is dominated by the $m=12$ component. $\delta\Phi_{m=12}(r)$ and $\delta{A}_{\|m=12}(r)$ have even parity along the radial direction with respect to $r=0.5$, which is different than the BAE for which $\delta{A}_{\|m=12}(r)$ has an odd parity. 

\begin{figure}
 \centering
        \includegraphics[width=0.4\textwidth]{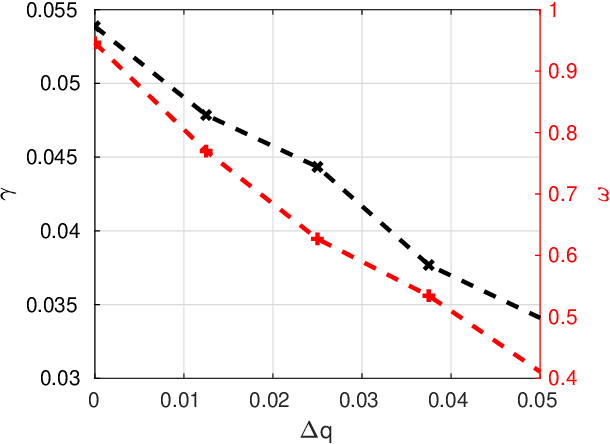}
        \includegraphics[width=0.37\textwidth]{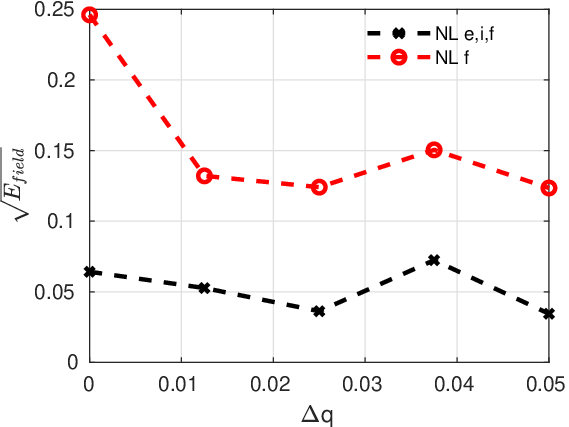}
 \caption{Left: RSAE growth rate (black line, left axis) and frequency (red line, right axis) for different $q$ profiles. Right: the saturation level of the RSAE with or without thermal plasma nonlinearity. }
\label{fig:saturationRSAE}
\end{figure}

The RSAE frequency is sensitive to the $q$ profile, which is characterized by a minimum $q_{\rm min}$ at $r_c=0.5a$, as shown in the left frame of Fig.~\ref{fig:saturationRSAE}.  The real frequency decreases significantly  as $\Delta q$ changes from $0$ to $0.05$, where $\Delta q\equiv q_{\rm min}-1.75$.  The growth rate shows a relatively weak variation, decreasing from $\sim0.05$ to $\sim0.035$ over the same range of $\Delta q$. The saturation level is shown in the right frame for cases with or without the thermal plasma nonlinearity. In all the cases, the saturation level is overestimated $2\sim4$ times when only the nonlinear EP model is considered, since the growth  rate and the corresponding saturation level remain similar across cases that include thermal plasma nonlinearity. It suggests that the overestimate of the saturation level in the absence of thermal plasma nonlinearity mainly depends on the growth rate, despite the frequency changes a lot. The thermal plasma nonlinearity leads to a lower RSAE saturation level and the corresponding suppressed stiffness of EP transport, which is beneficial for burning plasmas.

%%%%%%%%%%%%%%%%%%%%%%%%%%%%%%%%%%%%%%%%%%%%%%%%%%%%%%%%%%%%%%%%%%%%%%

%%%%%%%%%%%%%%%%%%%%%%%%%%%%%%%%%%%%%%%%%%%%%%%%%%%%%%%%%%%%%%%%%%%%%%

\section{Summary and outlook}
\label{sec:summary}
In this work, the influence of nonlinear thermal plasma dynamics on energetic particle (EP) transport has been investigated using global gyrokinetic simulations with the TRIMEG code for the ITPA TAE benchmark case and the extended BAE and RSAE cases. Two scenarios were compared: one in which only the energetic particles evolve nonlinearly, and another in which all plasma species are treated fully nonlinearly.

The linear properties of the instability are identical in both cases, indicating that the nonlinear dynamics of thermal ions and electrons do not affect the linear growth rate or mode frequency. However, clear differences emerge in the nonlinear phase. When the thermal plasma evolves nonlinearly, the fluctuation energy exhibits an overshoot and subsequently relaxes to a lower saturation level, whereas the case with only EP nonlinearity reaches a nearly constant saturation amplitude without a noticeable overshoot.

The nonlinear response of the thermal plasma significantly modifies the radial structure of the mode, leading to a broader fluctuation profile and changes in the dominant poloidal harmonics. These structural modifications have a direct impact on energetic particle transport. In particular, the dependence of the EP flux on the energetic particle drive becomes substantially less stiff when the nonlinear dynamics of thermal ions and electrons are included. While the case with EP-only nonlinearity exhibits an approximately quadratic scaling of the saturation level and EP flux, the fully nonlinear simulations show a scaling closer to linear.

These results demonstrate that nonlinear thermal plasma dynamics can play a key role in determining the saturation level of Alfvénic instabilities and the resulting energetic particle transport. Models that neglect nonlinear thermal plasma effects may therefore miss important feedback mechanisms relevant for realistic tokamak plasmas. In addition, the reduced stiffness of EP flux transport as the EP drive increases with the consideration of the thermal plasma nonlinearity, is consistent with the experimental observation that the EP transport flux saturates above a critical drive threshold \cite{collins2016observation}, even though the parameters and AE characteristics differ and the focus of the present work is distinct.  The avalanches and strong transport events are not strictly excluded. When conditions similar to a “dense” or “continuous” spectrum are met, such as for chorus \cite{zonca2022chorus,zonca2021chorus2} or a particularly strong anisotropic drive in the presence of a nearly periodic spectrum (e.g. energetic particle mode continuum similar to \cite{zonca2015nonlinear}), then energetic particle avalanches would still appear. Nevertheless, the results presented here provide additional insight into the factors that determine the saturation level of AEs and the resulting EP transport, regardless of whether multiple modes are present, as also emphasized in \cite{collins2016observation}.

Future work will focus on a detailed analysis of the underlying phase-space dynamics and on extending the present study to more realistic scenarios relevant for burning plasma experiments such as ITER, contributing to the development of advanced EP transport models \cite{carlevaro2022one,meng2024NF,lauber2024atep}. In addition, it merits more effort to develop realistic power source models and conduct long-time scale simulations to simulate the EP profile stiffness with respect to the beam power observed in the DIII-D experiment \cite{collins2016observation}. 

\ack{There are inspiring discussions between Z. Lu and Ningfei Chen about the theoretical interpretation of the TAE simulations. 
The simulations in this work were run on the local TOK cluster and the MPCDF Viper/Raven supercomputers. The EUROfusion projects TSVV-8, TSVV-10, TSVV-G, and ATEP are acknowledged. %Comments from Philipp Lauber are appreciated. 
This work has been carried out within the framework of the EUROfusion Consortium, funded by the European Union via the Euratom Research and Training Programme (Grant Agreement No 101052200—EUROfusion). Views and opinions expressed are, however, those of the author(s) only and do not necessarily reflect those of the European Union or the European Commission. Neither the European Union nor the European Commission can be held responsible for them.}

%\funding{Sample text inserted for demonstration.}
% This section is a list of funder names and grant numbers

%\roles{G. Meng: Simulations, code maintenance, analyses, writing, revision
%\\Z. Lu: Code development/maintenance, analyses, writing}
% List author names and the contributions made to the article, using terms from the NISO Contributor Roles Taxonomy (CRediT) https://credit.niso.org

\data{The data that support the findings of this study are included in the article and its supplementary material.}
% For more information on IOP Publishing's research data policy see: https://publishingsupport.iopscience.iop.org/questions/research-data/

\suppdata{Supplementary data are provided as MATLAB (.fig) files, containing the data and figure objects used to generate the results shown in the main text.}

%\section*{References}

\bibliographystyle{iopart-num}
\bibliography{reference}

\end{document}